\documentclass[aps,prd,letterpaper,showkeys,showpacs,twocolumn,nofootinbib,preprintnumbers,longbibliography,unsortedaddress]{revtex4-1}

\usepackage[top=2.8cm, bottom=2.8cm, left=2.4cm, right=2.4cm]{geometry}
\usepackage[T1]{fontenc}     
\usepackage{amsmath,amssymb,lmodern,amsfonts} 
\usepackage{graphicx}
\usepackage{hyperref}
\usepackage{color}
\usepackage{natbib}
\usepackage[caption=false]{subfig}
\usepackage{geometry}
\usepackage{booktabs}
\usepackage{amstext}

\makeatletter

\providecommand{\tabularnewline}{\\}

\makeatother

\newcommand{\lsim}{\lesssim}

\begin{document}

\title{Anisotropic solid dark energy}

\author{J. Motoa-Manzano}
\email{josue.motoa@correounivalle.edu.co}
\affiliation{Departamento  de  F\'isica,  Universidad  del Valle, \\ Ciudad  Universitaria Mel\'endez,  Santiago de Cali  760032,  Colombia}

\author{J. Bayron Orjuela-Quintana}
\email{john.orjuela@correounivalle.edu.co}
\affiliation{Departamento  de  F\'isica,  Universidad  del Valle, \\ Ciudad  Universitaria Mel\'endez,  Santiago de Cali  760032,  Colombia}

\author{Thiago S. Pereira}
\email{tspereira@uel.br}
\affiliation{Departamento de F\'isica, Universidade Estadual de Londrina, Rod. Celso Garcia Cid, Km 380, 86057-970, Londrina, Paran\'a, Brazil}

\author{C\'esar A. Valenzuela-Toledo}
\email{cesar.valenzuela@correounivalle.edu.co}
\affiliation{Departamento  de  F\'isica,  Universidad  del Valle, \\ Ciudad  Universitaria Mel\'endez,  Santiago de Cali  760032,  Colombia}

\begin{abstract}
In this paper, we study a triad of inhomogeneous scalar fields, known as ``solid'', as a source of homogeneous but anisotropic dark energy. By using a dynamical system approach, we find that anisotropic accelerated solutions can be realized as attractor points for suitable choices of the parameters of the model. We complement the dynamical analysis with a numerical solution whose initial conditions are set in the deep radiation epoch. The model can give an account of a non-negligible spatial shear within the observational bounds nowadays, even when the later is set to zero at early times. However, we find that there is a particular region in the parameter space of the model in which the Universe isotropizes. The anisotropic attractors, the particular isotropic region, and a nearly constant equation of state of dark energy very close to $- 1$ are key features of this scenario. Following a similar approach, we also analyzed the full isotropic version of the model. We find that the solid can be characterized by a constant equation of state and thus being able to simulate the behavior of a cosmological constant. 
\end{abstract}

\pacs{98.80.Cq; 95.36.+x}

\maketitle

\section{Introduction} \label{Introduction}
The current Universe is expanding at an accelerated rate. This fact was first discovered in type Ia supernovae (SNe Ia)  surveys \cite{Riess:1998cb, Schmidt:1998ys, Perlmutter:1998np} and later confirmed by several other observations like large scale structures (LSS) \cite{Tegmark:2003ud, Tegmark:2006az}, cosmic microwave background (CMB) \cite{Spergel:2003cb, Ade:2015xua}, and baryon acoustic oscillations (BAO)\cite{Percival:2007yw, Aubourg:2014yra} measurements. Observations also show that our expanding Universe is highly homogeneous, isotropic and spatially flat at cosmological scales \cite{deBernardis:2000sbo, Jaffe:2003it, Aghanim:2018eyx}. The simplest description of the Universe is based on the standard $\Lambda$ Cold Dark Matter ($\Lambda$CDM) model. In this model, the current accelerated expansion of the Universe is due to the repulsive effect of a constant energy density with negative pressure, which is given by the cosmological constant $\Lambda$ \cite{Amendola:2015ksp, Bamba:2012cp}. Despite its success, there are several theoretical and observational problems with this scenario. One of the theoretical difficulties is the so-called cosmological constant problem, which asserts that if $\Lambda$ is associated with the vacuum energy density of the Universe, the value predicted by the theory and the value obtained from observations differs by several tens orders of magnitude \cite{Weinberg:1988cp, Martin:2012bt}. On the observational side, the $H_0$ tension states that the current value of the Hubble parameter calculated from the CMB data does not agree with the value computed from local measurements of SNe Ia \cite{Riess:2016jrr, Riess:2019cxk}. It seems that this tension could be alleviated if extensions to $\Lambda$CDM model are considered \cite{DiValentino:2017iww, Guo:2018ans, Agrawal:2019lmo}.

The search for alternatives to the standard model is generally split into two broad categories: modified gravity theories and dynamical dark energy. While the former has been recently under observational pressure \cite{Collett:2018gpf, Ezquiaga:2018btd, He:2018oai, Do:2019txf, Abbott:2018lct}, the latter is usually based on time-dependent fields with vanishing spatial gradients \cite{Copeland:2006wr, Yoo:2012ug}. This choice ensures that the background geometry can be described by a homogeneous and isotropic metric, and thus the evolution of the Universe is also homogeneous and isotropic. On the other hand, some observations seem to imply a violation of the Universe's isotropy at large scales---the so called CMB anomalies \cite{Perivolaropoulos:2014lua, Schwarz:2015cma, Akrami:2019bkn}---suggesting that background metrics different to the homogeneous and isotropic Friedmann-Lema\^itre-Robertson-Walker (FLRW) metric should be considered \cite{Bennett:2010jb, Akrami:2019izv, Akrami:2019bkn}. It has been also pointed out that some of these CMB anomalies could be explained by the introduction of an anisotropic late-time accelerated expansion \cite{Battye:2009ze, Perivolaropoulos:2014lua, Schwarz:2015cma}. 

An anisotropic dark energy can be realized by considering a homogeneous but spatially anisotropic metric, together with a suitable arrangement of the fields driving the expansion of the universe. Among the proposals for anisotropic dark energy, we can find models of vector fields \cite{Koivisto:2008ig, Koivisto:2008xf, Thorsrud:2012mu}, p-forms \cite{Almeida:2019iqp, Guarnizo:2019mwf, Almeida:2018fwe} or non-Abelian gauge fields \cite{Orjuela-Quintana:2020klr, Guarnizo:2020pkj}. All of these models are based on time-dependent fields, so as to comply with the homogeneity of the background metric. Nonetheless, in Refs. \cite{ArmendarizPicon:2007nr, Endlich:2012pz}, it was shown that a triad of scalar fields with spatially constant but nonzero gradients can generate a homogeneous and isotropic energy-momentum tensor, i.e. invariant under translations and spatial rotations. This triad, known as ``solid'', is given by
\begin{equation} \label{ansatz}
\phi^I \equiv x^I, \quad I \in \{1, 2, 3\},
\end{equation}
where $\phi^I$ is a scalar field and $x^I$ is a comoving cartesian coordinate. The solid configuration for inhomogeneous scalar fields is similar to other configurations for different fields. For instance, the cosmic triad for vector and nonabelian gauge vector fields \cite{Bento:1992wy, ArmendarizPicon:2004pm, Golovnev:2008cf, Maleknejad:2011jw, Mehrabi:2017xga}, the U(1) triad of homogeneous scalar fields \cite{Firouzjahi:2018wlp}, and the recent Higgs triad \cite{Orjuela-Quintana:2020klr}, among others. 

In Refs. \cite{ArmendarizPicon:2007nr}, an isotropic energy-momentum tensor is achieved since the Lagrangians constructed from each of the three scalar fields are equal, while in Ref. \cite{Endlich:2012pz} it is assumed that the scalar fields possess an internal SO(3) symmetry such that the ansatz in Eq. \eqref{ansatz} is invariant under combined spatial and internal rotations. In Refs. \cite{Bartolo:2013msa, Bartolo:2014xfa}, it was shown that the solid configuration in Eq. \eqref{ansatz} can support \emph{prolonged anisotropic inflationary} solutions\footnote{Other scenarios of anisotropic inflation have been studied, see for instance \cite{Kanno:2008gn,Kanno:2008gn,Kanno:2008gn, Yokoyama:2008xw,Watanabe:2009ct,Watanabe:2010fh, Yamamoto:2012tq, Soda:2012zm,Ohashi:2013mka}.}. In this work, we show that this characteristic behavior is also present at late-times for most of the parameter space of the model  and thus the final stage of the Universe can be an anisotropic accelerated expansion, even if the initial spatial shear is set to zero. Nonetheless, there is a particular region in the parameter space of the model where the Universe becomes isotropic, meaning that the solid does not source the shear, which then eventually vanishes. For completeness, we also show that dark energy dominance is possible when the background metric is homogeneous and isotropic.

This paper is organized in the following way. In Sec. \ref{SE}, we present the action and the energy-momentum tensor of the model. In Sec. \ref{equations}, the equations of motion in an homogeneous but anisotropic background are derived. Section \ref{dynsys} is dedicated to the dynamical analysis of the model. A numerical integration of the background equations and the general cosmological evolution is presented in Sec. \ref{CE}. The isotropic version of the model is treated in Sec. \ref{Iso DE}. Finally, our conclusions are presented in Sec. \ref{conclusions}.

\section{General Model} \label{SE}

At this point, it is worth emphasizing that what we call a ``solid'' is the specific configuration of three inhomogeneous scalar fields given by Eq. \eqref{ansatz}.\footnote{Even more general configurations can be studied, like the ``supersolids''  in Ref. \cite{Celoria:2017bbh}}. However, different dynamics of the solid can be studied depending on the particular action constructed with the fields \eqref{ansatz}. For example, in Ref. \cite{Endlich:2012pz}, it is assumed that the solid itself is equipped with an internal SO(3) symmetry and its Lagrangian is a function of SO(3) invariants constructed from the matrix $B^{I J} \equiv \partial_\mu \phi^I \partial^\mu \phi^J$, being the Lagrangian compatible with a FLRW geometry. This same Lagrangian was studied in Refs. \cite{Bartolo:2013msa, Bartolo:2014xfa} but in an anisotropic background, where it was shown that prolonged anisotropic inflationary solutions can be obtained. Four our purposes, the assumption of an internal SO(3) symmetry is not necessary, and we thus opt for studying the simpler action
\begin{equation} \label{action}
S = \int \text{d}^4 x \sqrt{-g} \left[ \frac{m_\text{P}^2}{2} R - \sum_I F^I \left( X^I \right) + \mathcal{L}_m + \mathcal{L}_r \right],
\end{equation}
where $m_\text{P}$ is the reduced Planck mass, $R$ is the Ricci scalar, $F^I$ is the Lagrangian characterizing the dynamics of the scalar field $\phi^I$, whose argument is the canonical kinetic-type term
\begin{equation}
X^I \equiv g^{\mu \nu} \nabla_\mu \phi^I \nabla_\nu \phi^I,
\end{equation}
and $\mathcal{L}_m$ and $\mathcal{L}_r$ are the Lagrangians for matter and radiation fluids, respectively. This action is an extension to the late-time cosmology of the model studied in Ref. \cite{ArmendarizPicon:2007nr} in the inflationary context.

Varying the action in Eq. \eqref{action} with respect to the space-time metric $g^{\mu\nu}$, we get the gravitational field equations
\mbox{$m_\text{P}^2 G_{\mu\nu} = T_{\mu\nu}$}, with $G_{\mu\nu}$ the Einstein tensor and $T_{\mu\nu}$ the total energy tensor given by
\begin{equation} \label{energy tensor}
T_{\mu\nu} = 2 \sum_I F_{X^I} \nabla_\mu \phi^I \nabla_\nu \phi^I - g_{\mu\nu} \sum_I F^I + T^m_{\mu\nu} + T^r_{\mu\nu}, 
\end{equation}
where $T^m_{\mu\nu}$ and $T^r_{\mu\nu}$ are the energy tensors associated to the matter and radiation perfect fluids, respectively, and we have used the shorthand notation \mbox{$F_{X^I} \equiv \frac{\text{d} F^I}{\text{d} X^I}$}. Varying the action with respect to $\phi^I$, we get the equation of motion
\begin{equation}
\nabla_\mu \left( F_{X^I} \nabla^\mu \phi^I \right) = 0.
\end{equation}

\section{Background Equations of Motion} \label{equations}

Since we are interested in anisotropic deformations sourced by the solid, we adopt the geometry of a  homogeneous but anisotropic Bianchi-I metric. For simplicity, we assume that there exists a residual isotropy in the $(y, z)$ plane, such that the background geometry is given by
\begin{equation} \label{metric}
\text{d} s^2 = - \text{d} t^2 + a(t)^2 \left[ e^{-4 \sigma (t)} \text{d} x^2 + e^{2 \sigma (t)} \left( \text{d} y^2 + \text{d} z^2 \right) \right],
\end{equation}
where $a(t)$ is the average scale factor and $\sigma(t)$ is the geometrical shear, being both functions of the cosmic time $t$. 
In Ref. \cite{ArmendarizPicon:2007nr}, it was shown that the action in Eq. \eqref{action} can be compatible with the symmetries of the FLRW metric if the three Lagrangians $F^I$ are identical; i.e. $F^I = F$ and thus $\sum_I F^I = 3F$.  However, in the Bianchi-I background in Eq. \eqref{metric}, the canonical kinetic-type terms read
\begin{equation}
X^1(t) = \frac{e^{4 \sigma(t)}}{a^2(t)}, \quad X^2(t) = X^3 (t) = \frac{e^{- 2 \sigma(t)}}{a^2(t)},
\end{equation}
and the requirement for the Lagrangians $F^I$ in this case is
\begin{equation} 
\sum_I F^I \left( X^I \right) = F^1 \left( X^1 \right) + 2 F^2 \left( X^2 \right).
\end{equation} 

Considering the ``$00$'' component of the gravitational field equations, the first ``Friedman'' equation reads
\begin{equation} \label{H2 eq}
3 m_\text{P}^2 H^2 = F^1 + 2 F^2 + 3 m_\text{P}^2 \dot{\sigma}^2 + \rho_m + \rho_r,
\end{equation}
where $H = \dot{a} / a$ is the Hubble parameter,\footnote{Here, an overdot denotes a derivative with respect to the cosmic time $t$.} and we have defined $\rho_m$ and $\rho_r$ as the densities for the matter and radiation perfect fluids, respectively. The second Friedman equation follows from \mbox{$m_\text{P}^2 \text{tr} \, G_{\mu\nu} = \text{tr} \, T_{\mu\nu}$}, which can be written as
\begin{equation} \label{Hdot eq}
- 2 m_\text{P}^2 \dot{H} = \frac{2}{3} X^1 F_{X^1} + \frac{4}{3} X^2 F_{X^2} + \rho_m + \frac{4}{3} \rho_r + 6 m_\text{P}^2 \dot{\sigma}^2.
\end{equation}
Finally, the evolution equation for the geometrical shear is obtained from the relation \mbox{$m_\text{P}^2(G^2_{\ 2} - G^1_{\ 1}) = T^2_{\ 2} - T^1_{\ 1}$} as
\begin{equation} \label{sigma eq}
\ddot{\sigma} + 3 H \dot{\sigma} = \frac{2}{3 m_\text{P}^2} \left( X^2 F_{X^2} - X^1 F_{X^1} \right).
\end{equation}
Since we are interested in anisotropic late-time accelerated solutions, it is necessary to characterize the dark energy fluid. We define the density and pressure of dark energy by
\begin{align*}
\rho_\text{DE} &\equiv F^1 + 2 F^2 + 3 m_\text{P}^2 \dot{\sigma}^2, \\
p_\text{DE} &\equiv \frac{1}{3} (2 x_1 - 3) F^1 + \frac{2}{3} (2 x_2 - 3) F^2 + 3 m_\text{P}^2 \dot{\sigma}^2,
\end{align*}
where we have defined the quantities
\begin{equation}
x_1 \equiv \frac{X^1 F_{X^1}}{F^1}, \quad x_2 \equiv \frac{X^2 F_{X^2}}{F^2},
\end{equation}
which characterize the form of the Lagrangians $F^1$ and $F^2$, respectively.

Note that our choice to include the geometrical shear in $\rho_\text{DE}$ and $p_\text{DE}$ (instead of considering only the contribution coming from the solid) allows us to write the continuity equation simply as
\begin{equation}
\dot{\rho}_\text{DE} + 3 H \left( \rho_\text{DE} + p_\text{DE} \right) = 0,
\end{equation}
which greatly simplifies our analysis.\footnote{Had we chosen to separate the contributions of the solid and the geometry, we would end up with an equation of the form $\dot{\bar{\rho}}_\text{DE} + 3 H \left(\bar{\rho}_\text{DE} + \bar{p}_\text{DE} \right) \propto \dot{g}_{ij}\pi^{ij}$, where $\dot{g}_{ij}$ is the time derivative of the spatial part of the metric, and $\pi^{ij}$ is the trace-free part of the energy-momentum tensor.}

The set of Eqs. \eqref{H2 eq}-\eqref{sigma eq} describes the cosmological background dynamics. In the next section, we will study the asymptotic behavior of this set of equations through a dynamical system analysis \cite{Wainwright2009, Coley:2003mj}.

\section{Dynamical system} \label{dynsys}

\subsection{Autonomous System}

In order to proceed, we introduce the following dimensionless variables
\begin{equation*}
f_1^2 \equiv \frac{F^1}{3 m_\text{P}^2 H^2}, \quad f_2^2 \equiv \frac{F^2}{3 m_\text{P}^2 H^2}, \quad \Omega_m \equiv \frac{\rho_m}{3 m_\text{P}^2 H^2}, 
\end{equation*}
\begin{equation} \label{variables}
\Omega_r \equiv \frac{\rho_r}{3 m_\text{P}^2 H^2}, \quad \Sigma \equiv \frac{\dot{\sigma}}{H}, \\
\end{equation}
such that the first Friedman equation \eqref{H2 eq} becomes the constraint
\begin{equation}\label{constraint}
\Omega_m = 1 - f_1^2 - 2 f_2^2 - \Sigma^2 - \Omega_r\,.
\end{equation}
Changing the cosmic time $t$ for the number $N$ of $e$-folds defined as 
$\text{d} N \equiv H \text{d} t$, the background equations \eqref{H2 eq}-\eqref{sigma eq} are replaced by the autonomous system\footnote{Here, a prime denotes a derivative with respect to the number of $e$-folds $N$.}
\begin{align}
f_1' &= f_1 \left[ q + 1 - x_1 \left(1 - 2 \Sigma \right) \right], \label{f1 eq}\\
f_2' &= f_2\left[ q + 1 - x_2 \left(1 + \Sigma \right) \right], \\
\Sigma' &= \Sigma (q - 2) - 2 \left( x_1 f_1^2 - x_2 f_2^2 \right), \\
\Omega_r' &= 2 \Omega_r \left( q - 1 \right), \label{r eq}
\end{align}
where the deceleration parameter, $q \equiv - a \ddot{a} / \dot{a}^2$, is given by
\begin{equation}
q = \frac{1}{2} \left[ 1 + (2 x_1 - 3) f_1^2 + 2 (2 x_2 - 3) f_2^2 + 3 \Sigma^2 + \Omega_r \right].
\end{equation}
However, instead of the deceleration parameter, we equivalently characterize the evolution of the average scale factor $a(t)$ in terms of the effective equation of state $w_\text{eff} \equiv (2 q - 1) / 3$.

The dark sector is characterized by its equation of state $w_\text{DE} \equiv p_\text{DE} / \rho_\text{DE}$, which in terms of the dynamical variables reads
\begin{equation}
w_\text{DE} = - 1 + \frac{2}{3} \, \frac{x_1 f_1^2 + 2 x_2 f_2^2 + 3 \Sigma^2}{f_1^2 + 2 f_2^2 + \Sigma^2},
\end{equation}
and its density parameter $\Omega_\text{DE} \equiv \rho_\text{DE} / 3 m_\text{P}^2 H^2$.
 
Since the functions $x_1$ and $x_2$ cannot themselves be expressed in terms of the dimensionless variables, it is necessary to choose the specific Lagrangians $F^1$ and $F^2$ in order to get a closed autonomous system. In this case, the simplest model is obtained when $x_1$ and $x_2$ are constants, which corresponds to a power law model
\begin{equation*}
F^1 \propto (X^1)^n, \quad F^2 \propto (X^2)^m,  
\end{equation*}
such that
\begin{equation}
x_1 = n, \quad x_2 = m.
\end{equation}
We want to stress that a different choice would yield to time-dependent parameters $x_1$ and $x_2$, such that, in principle, the equation of state of dark energy could vary in unimagined ways. Due to the lacking of restrictions in the functional form of the Lagrangians $F^I$ (some of them could be obtained from a reconstruction method, for example), we focus in this particular choice by its simplicity. In the next subsection, we will study the asymptotic behavior of the system by finding the fixed points of the autonomous system, and we will see that this simple choice is enough to get interesting behaviors.

\subsection{Fixed Points and Stability}

In the following, we discuss the fixed points relevant to the radiation ($\Omega_r \simeq 1, w_\text{eff} \simeq 1/3$), matter ($\Omega_m \simeq 1, w_\text{eff} \simeq 0$), and dark energy eras ($\Omega_\text{DE} \simeq 1, w_\text{eff} < -1/3$), which can be obtained by setting $f_1' = 0$, $f_2' = 0$, $\Sigma' = 0$, and $\Omega_r' = 0$ in equations \eqref{f1 eq}-\eqref{r eq}. The stability of these points can be known by perturbing the autonomous set around them. Up to linear order, the perturbations \mbox{$\delta \mathcal{X} = \left( \delta f_1, \delta f_2, \delta \Sigma, \delta \Omega_r \right)$} satisfy the differential equation,
\begin{equation}
\delta \mathcal{X}' = \mathbb{M} \, \mathcal{X}, 
\end{equation} 
where $\mathbb{M}$ is a $4 \times 4$ Jacobian matrix. The sign of the real part of the eigenvalues $\lambda_{1, 2, 3, 4}$ of $\mathbb{M}$ determines the stability of the point. A fixed point is an attractor, or sink, if the real part of all the eigenvalues are negative. If at least one of the eigenvalues has positive real part  
it is called a saddle. If the real part of all the eigenvalues are positive the fixed point is called a repeller or source. 

In what follows, we refer to each point by its name, which is defined as $R$, $M$ or $DE$ -- depending on wheter it corresponds to a radiation, matter or dark energy dominated universe -- followed by a number. The points and their eigenvalues are gathered in Tables~\ref{table-fp} and~\ref{table-ev}, respectively.

\begin{center}
\begin{table*}
\begin{centering}
\begin{tabular}{cccccccc}
\toprule 
Fixed Point & $f_{1}$ & $f_{2}$ & $\Omega_{r}$ & $\Omega_{m}$ & $\Sigma$ & $w_{\text{eff}}$ & stability\tabularnewline
\midrule
\midrule 
\emph{R-1} & 0 & 0 & 1 & 0 & 0 & 1/3 & saddle\tabularnewline
\emph{R-2} & $\frac{\sqrt{2-n}}{2n}$ & 0 & $\frac{3n^{2}+5n-6}{4n^{2}}$ & 0 & $\frac{n-2}{2n}$ & 1/3 & saddle\tabularnewline
\emph{R-3} & 0 & $\frac{\sqrt{2-m}}{\sqrt{2}m}$ & $\frac{5m-6}{m^{2}}$ & 0 & $\frac{2-m}{m}$ & 1/3 & saddle\tabularnewline
\emph{M-1} & 0 & 0 & 0 & 1 & 0 & 0 & saddle/attractor\tabularnewline
\emph{M-2} & $\frac{\sqrt{3(3-2n)}}{4n}$ & 0 & 0 & $\frac{3(2n^{2}+3n-3)}{8n^{2}}$ & $\frac{2n-3}{4n}$ & 0 & saddle/attractor\tabularnewline
\emph{M-3} & 0 & $\frac{\sqrt{3(3-2m)}}{\sqrt{8}m}$ & 0 & $\frac{9(m-1)}{2m^{2}}$ & $\frac{3-2m}{2m}$ & 0 & saddle/attractor\tabularnewline
\emph{DE-1} & $\frac{\sqrt{3(3+n)(1-n)}}{3-n}$ & 0 & 0 & $0$ & $\frac{2n}{n-3}$ & $-1+\frac{2n(1+n)}{3-n}$ & saddle/attractor\tabularnewline
\emph{DE-2} & 0 & $\frac{\sqrt{3(3/2-m)}}{3-m}$ & 0 & 0 & $\frac{m}{3-m}$ & $-1+\frac{2m}{3-m}$ & saddle/attractor\tabularnewline
\emph{DE-3} & $\frac{\sqrt{3(m-n+mn)}}{m+2n}$ & $\frac{\sqrt{3\left[n(n+1)+m(n-1)\right]}}{\sqrt{2}(m+2n)}$ & 0 & 0 & $\frac{n-m}{m+2n}$ & $-1+\frac{2mn}{m+2n}$ & saddle/attractor\tabularnewline
\bottomrule
\end{tabular}
\par\end{centering}
\caption{Fixed points for the dynamical system \eqref{f1 eq}-\eqref{r eq}. The points are labelled according to the cosmological regime as \emph{R-} (radiation), \emph{M}- (matter) and \emph{DE-} (dark energy).}
\label{table-fp}
\end{table*}
\par\end{center}

\begin{center}
\begin{table*}
\begin{centering}
\begin{tabular}{ccccc}
\toprule 
Fixed Point & $\lambda_{1}$ & $\lambda_{2}$ & $\lambda_{3}$ & $\lambda_{4}$\tabularnewline
\midrule
\midrule 
\emph{R-1} & $-1$ & $1$ & $2-m$ & $2-n$\tabularnewline
\emph{R-2} & $1$ & $2\left[1+\frac{m}{2}\left(\frac{1}{n}-\frac{3}{2}\right)\right]$ & $-\frac{1}{2}\left[1-\frac{\sqrt{6n^{3}-n^{2}-32n+24}}{n}\right]$ & $-\frac{1}{2}\left[1+\frac{\sqrt{6n^{3}-n^{2}-32n+24}}{n}\right]$\tabularnewline
\emph{R-3} & $1$ & $2\left[1+\frac{n}{2}\left(\frac{4}{m}-3\right)\right]$ & $-\frac{1}{2}\left[1-\frac{\sqrt{41m^{2}-128m+96}}{m}\right]$ & $-\frac{1}{2}\left[1+\frac{\sqrt{41m^{2}-128m+96}}{m}\right]$\tabularnewline
\emph{M-1} & $-\frac{3}{2}$ & $-1$ & $\frac{3}{2}-m$ & $\frac{3}{2}-n$\tabularnewline
\emph{M-2} & $-1$ & $\frac{3}{2}\left[1-m\left(1-\frac{1}{2n}\right)\right]$ & $-\frac{3}{4}\left[1-\frac{\sqrt{(n^{2}+n-3)(4n-3)}}{n}\right]$ & $-\frac{3}{4}\left[1+\frac{\sqrt{(n^{2}+n-3)(4n-3)}}{n}\right]$\tabularnewline
\emph{M-3} & $-1$ & $\frac{3}{2}\left[1+2n\left(\frac{1}{m}-1\right)\right]$ & $-\frac{3}{4}\left[1-\frac{\left(6-5m\right)}{m}\right]$ & $-\frac{3}{4}\left[1+\frac{\left(6-5m\right)}{m}\right]$\tabularnewline
\emph{DE-1} & $\frac{3[m(n-1)+n(n+1)]}{3-n}$ & $\frac{2(3n^{2}+5n-6)}{3-n}$ & $\frac{6n^{2}+9n-9}{3-n}$ & $\frac{3(n^{2}+2n-3)}{3-n}$\tabularnewline
\emph{DE-2} & $\frac{2(5m-6)}{3-m}$ & $9\left(\frac{m-1}{3-m}\right)$ & $3\left(\frac{2m-3}{3-m}\right)$ & $\frac{3[m+n(m-1)]}{3-m}$\tabularnewline
\emph{DE-3} & \multicolumn{4}{c}{too long to show}\tabularnewline
\bottomrule
\end{tabular}
\par\end{centering}
\caption{Eigenvalues for the equilibrium points in Table~\ref{table-fp}. The expressions for the eigenvalues of the point \emph{DE-3} are too long, and thus omitted.}\label{table-ev}
\end{table*}
\par\end{center}


\subsubsection{\textbf{Radiation Dominance}}

\paragraph{$\bullet$ ($R$-1) Isotropic radiation:}

This point corresponds to an isotropic radiation-dominated universe, and it trivially satisfies the constraint \eqref{constraint}. One can check that the eigenvector associated with the eigenvalue ${\lambda_1 = -1}$ points in the $\Sigma$ direction in the phase space $(f_1, f_2, \Sigma, \Omega_r)$, indicating that the trajectories around this point are attracted in this direction. This means that the shear decays from its value around this point. On the other hand, the eigenvector associated with the eigenvalue $\lambda_2 = 1$ points to the $\Omega_r$ direction, meaning that radiation is running away from its value $\Omega_r = 1$. The eigenvalues $\lambda_3$ and $\lambda_4$ are positive for $m$ and $n$ less than 2, respectively. Under this condition, (\emph{R-1}) is a saddle with three positive eigenvalues, and the dark components $f_1$ and $f_2$ grow during the radiation epoch, since the eigenvectors associated to these eigenvalues ($\lambda_{3, 4}$) point to the $f_2$ and $f_1$ directions, respectively. \\

\paragraph{$\bullet$ ($R$-2) Anisotropic radiation scaling with $F^1$:}
This corresponds to an anisotropic solution where the density parameter and equation of state of dark energy are given by
\begin{equation}
\Omega_{\text{DE}} = \frac{(n - 2)(n - 3)}{4 n^2}, \quad w_{\text{DE}} = \frac{1}{3},
\end{equation}
indicating that dark energy scales as a radiation fluid, or ``dark radiation''. This point is a viable solution if the conditions
\[
f_1^2 \geq 0\,,\; 0 \leq \Omega_\text{DE} \leq 1\,,\quad \text{and}\quad 
0 \leq \Omega_r \leq 1,
\]
are satisfied. Furthermore, the big-bang nucleosynthesis (BBN) gives the bound $\Omega_{\text{DE}} < 0.045$ \cite{Bean:2001wt}. Imposing all of these conditions, we determine that the physical region of existence of the point (\emph{R-2}) is
\begin{equation} \label{R1 region}
1.64237 < n \leq 2, \quad \forall \quad m.
\end{equation}

The eigenvalues of $\mathbb{M}$ in this point (see Table~\ref{table-ev}) tell us that this is a saddle point. The eigenvalues $\lambda_3$ and $\lambda_4$ are negative in the region of existence given in Eq. \eqref{R1 region}. The second eigenvalue is positive in the region of existence of the point for $m < 4n / (3n - 2)$, and negative otherwise. Since the eigenvector associated to this eigenvalue points to the $f_2$ direction when $\lambda_2 > 0$, the dark component $f_2$ grows during the radiation epoch. \\

\paragraph{$\bullet$ ($R$-3) Anisotropic radiation scaling with $F^2$:}
The density parameter and equation of state of dark energy in this solution are given by
\begin{equation}
\Omega_{\text{DE}} = \frac{(m - 2)(m - 3)}{m^2}, \quad w_{\text{DE}} = \frac{1}{3}\,.
\end{equation}
Imposing the conditions 
\[
f_2^2 \geq 0\,,\quad 0 \leq \Omega_\text{DE} \leq 1\,,\quad \text{and} \quad
0 \leq \Omega_r \leq 1,
\]
as well as the BBN bound $\Omega_{\text{DE}} < 0.045$, the physical region of existence of (\emph{R-3}) becomes
\begin{equation} \label{R2 region}
1.86271 < m \leq 2, \quad \forall \quad n\,.
\end{equation}
The eigenvalues of $\mathbb{M}$ in this point show that this is a saddle point. The eigenvalues $\lambda_3$ and $\lambda_4$ are negative in the region of existence given in Eq. \eqref{R2 region}. The second eigenvalue is positive in the region of existence of the point for $n \leq 2m / (3m - 4)$, and negative otherwise. Since the eigenvector associated to this eigenvalue points to the $f_1$ direction, when $\lambda_2 > 0$, the dark component $f_1$ grows during the radiation epoch.

\subsubsection{\textbf{Matter Dominance}}

\paragraph{$\bullet$ ($M$-1) Isotropic matter:}
This corresponds to an isotropic matter-dominated universe with 
$\Omega_\text{DE} = 0$, and $w_\text{DE}$ undetermined. The eigenvector associated with the eigenvalue $\lambda_1 = -3 / 2$ points to the $\Sigma$ direction indicating that the  trajectories around this point are attracted in this direction (faster than around the point (\emph{R-1})). In this case, the eigenvector associated with the eigenvalues $\lambda_2 = - 1$ points to the $\Omega_r$ direction, meaning that radiation is decaying. The eigenvalues $\lambda_3$ and $\lambda_4$ are positive for $m$ and $n$ less than $3/2$, respectively. Under this condition, (\emph{M-1}) is a saddle with two positive eigenvalues, and the dark components $f_1$ and $f_2$ grow during the matter epoch, since the eigenvector associated to these eigenvalues point to the $f_2$ and $f_1$ directions, respectively. \\

\paragraph{$\bullet$ ($M$-2) Anisotropic matter scaling with $F^1$:}
The energy-density and equation of state of dark energy in this solution are given by
\begin{equation}
\Omega_{\text{DE}} = \frac{(2n - 3)(n - 3)}{8 n^2}, \quad w_{\text{DE}} = 0,
\end{equation}
indicating that $\rho_{\text{DE}}$ scales as a pressureless fluid. This point is a viable solution if 
\[
f_1^2 \geq 0\,,\quad 0 \leq \Omega_\text{DE} \leq 1\,, \quad \text{and} \quad
0 \leq \Omega_m \leq 1\,,
\]
are satisfied. Moreover, CMB anisotropies give the bound $\Omega_{\text{DE}} < 0.02$ around the redshift $z = 50$ \cite{Aghanim:2018eyx} (which ensures that we are deep in the matter-dominated era). Therefore, the physical region of existence of (\emph{M-2}) is 
\begin{equation} \label{M1 region}
1.40166 < n \leq 1.5, \quad \forall \quad m\,.
\end{equation}
The eigenvalues $\lambda_3$ and $\lambda_4$ are negative in the region of existence given in Eq. \eqref{M1 region}. For this point to be a possible candidate for the matter-dominated epoch, it has to be a saddle rather than an attractor in order to allow a subsequent accelerated expansion epoch. This means that the second eigenvalue has to be positive. We have $\lambda_2 > 0$ in the region of existence of the point for \mbox{$m < 2n / (2n - 1)$}. Since the eigenvector associated to this eigenvalue points to the $f_2$ direction, when $\lambda_2 > 0$, the dark component $f_2$ grows during the matter epoch. \\

\paragraph{$\bullet$ ($M$-3) Anisotropic matter scaling with $F^2$:}
For this point, the energy-density and equation of dark energy are
\begin{equation}
\Omega_{\text{DE}} = \frac{(2m - 3)(m - 3)}{2 m^2}, \quad w_{\text{DE}} = 0\,.
\end{equation}
Imposing the conditions 
\[
f_2^2 \geq 0\,,\quad 0 \leq \Omega_\text{DE} \leq 1\,,\quad \text{and} \quad 
0 \leq \Omega_r \leq 1,
\]
together with the CMB bound $\Omega_{\text{DE}} < 0.02$, the physical region 
of existence of (\emph{M-2}) is found to be
\begin{equation} \label{M2 region}
1.47166 < m \leq 1.5, \quad \forall \quad n\,.
\end{equation}
The eigenvalues $\lambda_3$ and $\lambda_4$ are negative in the region of existence given in Eq. \eqref{M2 region}. For this point to be a saddle, the second eigenvalue has to be positive, which is the case in its region of existence if $n \leq m / (2m - 2)$. Since the eigenvector associated to this eigenvalue points to the $f_1$ direction, when $\lambda_2 > 0$, the dark component $f_1$ grows during the matter epoch. 

\subsubsection{\textbf{Dark Energy Dominance}}

\paragraph{$\bullet$ ($DE$-1) Anisotropic dark energy scaling with $F^1$:}
This is the first solution corresponding to an anisotropic dark energy dominated universe. The energy density and equation of state of dark energy 
are given by
\begin{equation} \label{FixedDE1wDE}
\Omega_{\text{DE}} = 1, \quad w_{\text{DE}} = w_\text{eff} = -1 + \frac{2n(1 + n)}{3 - n}\,.
\end{equation}
If we now impose the conditions $f_1^2 \geq 0$ and \mbox{$-1 \leq w_\text{DE} < - 1/3$} -- the latter being necessary for having accelerated solutions\footnote{We concentrate in no-ghost solutions; i.e. $w_\text{DE} \geq -1$, although this possibility has not been discarded by observations yet \cite{Aghanim:2018eyx}.} -- we find the following regions of existence\footnote{The symbol $\lor$ stands for the logic ``OR''.}
\begin{equation}\label{DE1-2regions}
- 1.86852 < n \leq -1 \quad \lor \quad 0 \leq n < 0.535184\,.
\end{equation} 
However, since current observations favour an equation of state of dark energy $w_\text{DE} \approx - 1$ nowadays \cite{Aghanim:2018eyx}, \emph{and} a small anisotropy\footnote{Here, the subscript $0$ means that the corresponding quantity is evaluated nowadays.} $| \Sigma_0 | < \mathcal{O}(0.001)$ \cite{Campanelli:2010zx, Amirhashchi:2018nxl}, we take the region of existence of this point as
\begin{equation} \label{DE1 region}
0 \leq n < 0.535184, \quad \forall \quad m\,.
\end{equation}
The first branch in \eqref{DE1-2regions}, although leading to a viable equation of state for dark energy, leads to a too large $|\Sigma_0|$, and is thus discarded. Note that if $n = 0$, then $\Sigma = 0$ given that the Lagrangian $F^1$ becomes a  cosmological constant.

The eigenvalues $\lambda_{2, 3, 4}$ are negative in the region of existence given in Eq. \eqref{DE1 region}, while $\lambda_1$ is negative in this region when
\begin{equation} \label{DE1 attractor}
m \geq n \left( \frac{1 + n}{1 - n} \right).
\end{equation}
Therefore, we conclude that (\emph{DE-1}) is an attractor inside the region of existence in Eq. \eqref{DE1 region} whenever Eq. \eqref{DE1 attractor} is obeyed. \\
 
\paragraph{$\bullet$ ($DE$-2) Anisotropic dark energy with $F^2$:}
The dark energy parameters in this solution are
\begin{equation} \label{DE2 wDE}
\Omega_{\text{DE}} = 1, \quad w_{\text{DE}} = w_\text{eff} = -1 + \frac{2 m}{3 - m}\,.
\end{equation}
Imposing the conditions $f_2^2 \geq 0$ and \mbox{$-1 \leq w_\text{DE} < - 1/3$}, we arrive at the following region of existence
\begin{equation} \label{DE2 region}
0 \leq m < 0.75, \quad \forall \quad n.
\end{equation} 
Note that if $m = 0$, then $\Sigma = 0$ given that the Lagrangian $F^2$ becomes a  cosmological constant.

The eigenvalues $\lambda_{1, 2, 3}$ are negative in the region of existence given by Eq. \eqref{DE2 region}, while $\lambda_4$ is negative in this region when
\begin{equation} \label{DE2 attractor}
n \geq \frac{m}{1 - m}\,.
\end{equation}
Thus, (\emph{DE-2}) is an attractor inside the region of existence in Eq. \eqref{DE2 region} when Eq. \eqref{DE2 attractor} is satisfied. \\

\begin{figure}[t!]
\includegraphics[width=0.92\linewidth]{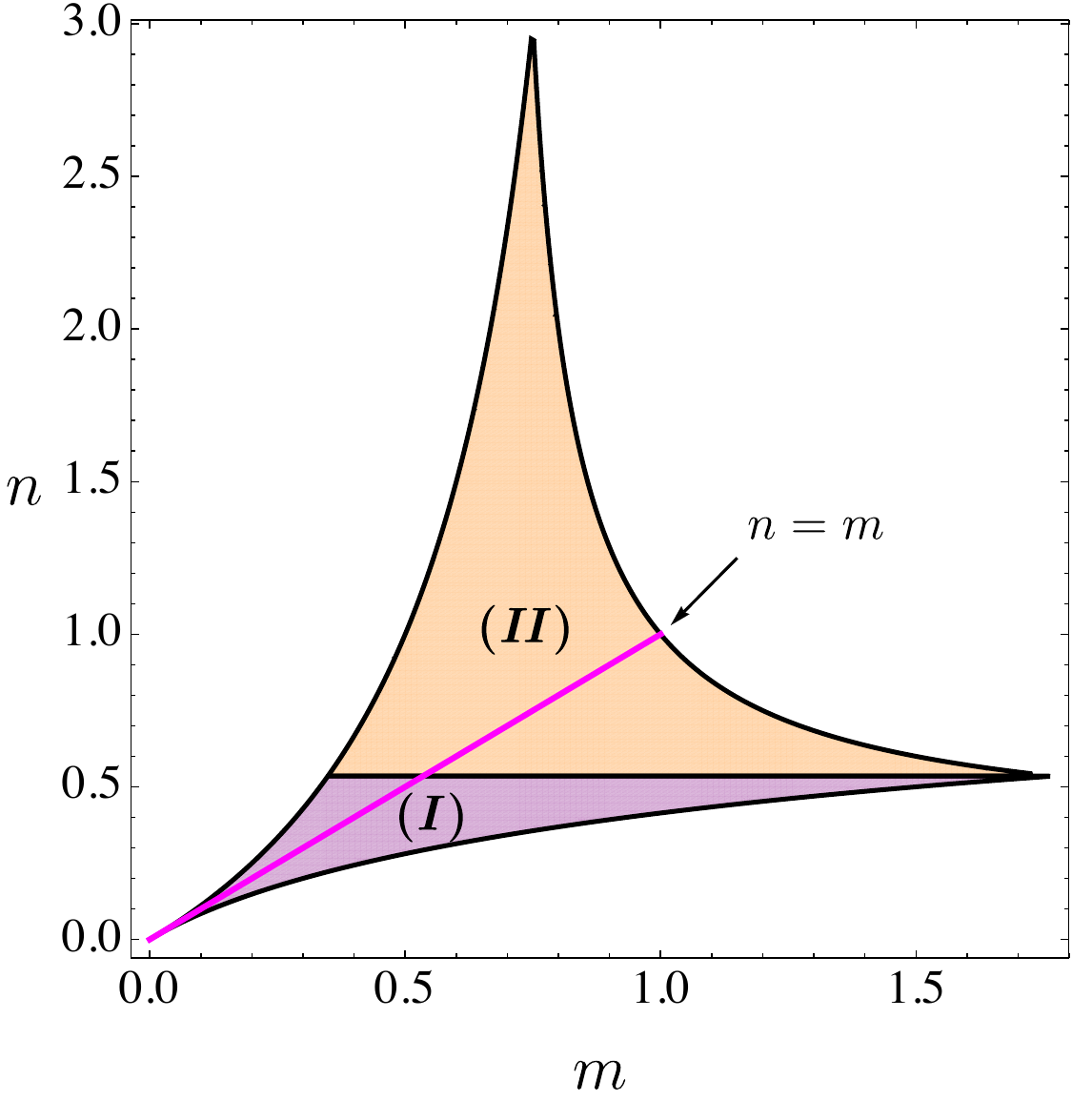}    
\caption{Region of existence of (\emph{DE-3}) when it is an accelerated solution without phantom line crossing. Regions (\emph{I}) and (\emph{II}) are given by Eqs. \eqref{Region I} and \eqref{Region II}, respectively. The magenta line represents the points where $n = m$, i.e. the points where the shear is zero.}
\label{RegionsDE3}
\end{figure}

\paragraph{$\bullet$ ($DE$-3) Anisotropic dark energy scaling with $F^1$ and $F^2$:} \label{SecDE3}
This is the only solution in which dark energy scales with both $F^1$ and $F^2$. The parameters of dark energy in this case are
\begin{equation} \label{FixedDE3wDE}
\Omega_{\text{DE}} = 1, \quad w_{\text{DE}} = w_\text{eff} = -1 + \frac{2 m n}{m + 2n}\,.
\end{equation}
By demanding that both $f_1^2$ and $f_2^2$ are positive, and that \mbox{$-1 \leq w_\text{DE} < - 1/3$}, we arrive at two possible regions for the parameters $m$ and $n$:\footnote{The symbol $\land$ stands for the logic ``AND''.}
\begin{align}
(I)&: 0 < n < 0.535184 \ \land \ \frac{n}{1 + n} \leq m < n \left( \frac{1 + n}{1 - n} \right), \label{Region I} \\
(II)&: 0.535184 \leq n < 3 \ \land \ \frac{n}{1 + n} \leq m < \frac{2n}{3n - 1}\,. \label{Region II}
\end{align} 
These regions are plotted in Fig.~\ref{RegionsDE3}. Notice that the cases $m = 0$ and $n = 0$ are not allowed in the region of existence. 
Note that in the region of existence of ($DE$-3) there is the possibility that $n = m$ (magenta line in Fig. \ref{RegionsDE3}), implying that $\Sigma = 0$, as it can be seen in Table \ref{table-fp}. This follows since, in the case $n = m$, $f_1 = f_2$ (see Table \ref{table-fp}) and therefore $F^1 = F^2$ by the definition of the variables. This implies that the right-hand side of Eq. \eqref{sigma eq} is zero, and thus the shear decays since the solid is not sourcing it. In other words, the shear is dynamically erased given that the Lagrangians behave in the same way. On the other hand, and as we will see later, the shear grows at late-times even if it is set to zero as the initial condition, since the solid sources it given that the Lagrangians $F^I$ behave in different ways, i.e. when $n \neq m$.

The eigenvalues of $\mathbb{M}$ in this fixed point are given by too long algebraic expressions involving the parameters $n$ and $m$. We omit them here since only the sign of the real part of the eigenvalues is relevant to the stability analysis. We investigated the parameter window, inside the regions of existence in Fig. \ref{RegionsDE3}, where $\lambda_{1, 2, 3, 4}$ are negative, and we found that the four eigenvalues are negative inside the whole region of existence of the point; i.e. when (\emph{DE-3}) exists, it \emph{is} an attractor. \\

\begin{figure}[t!]
\includegraphics[width=0.92\linewidth]{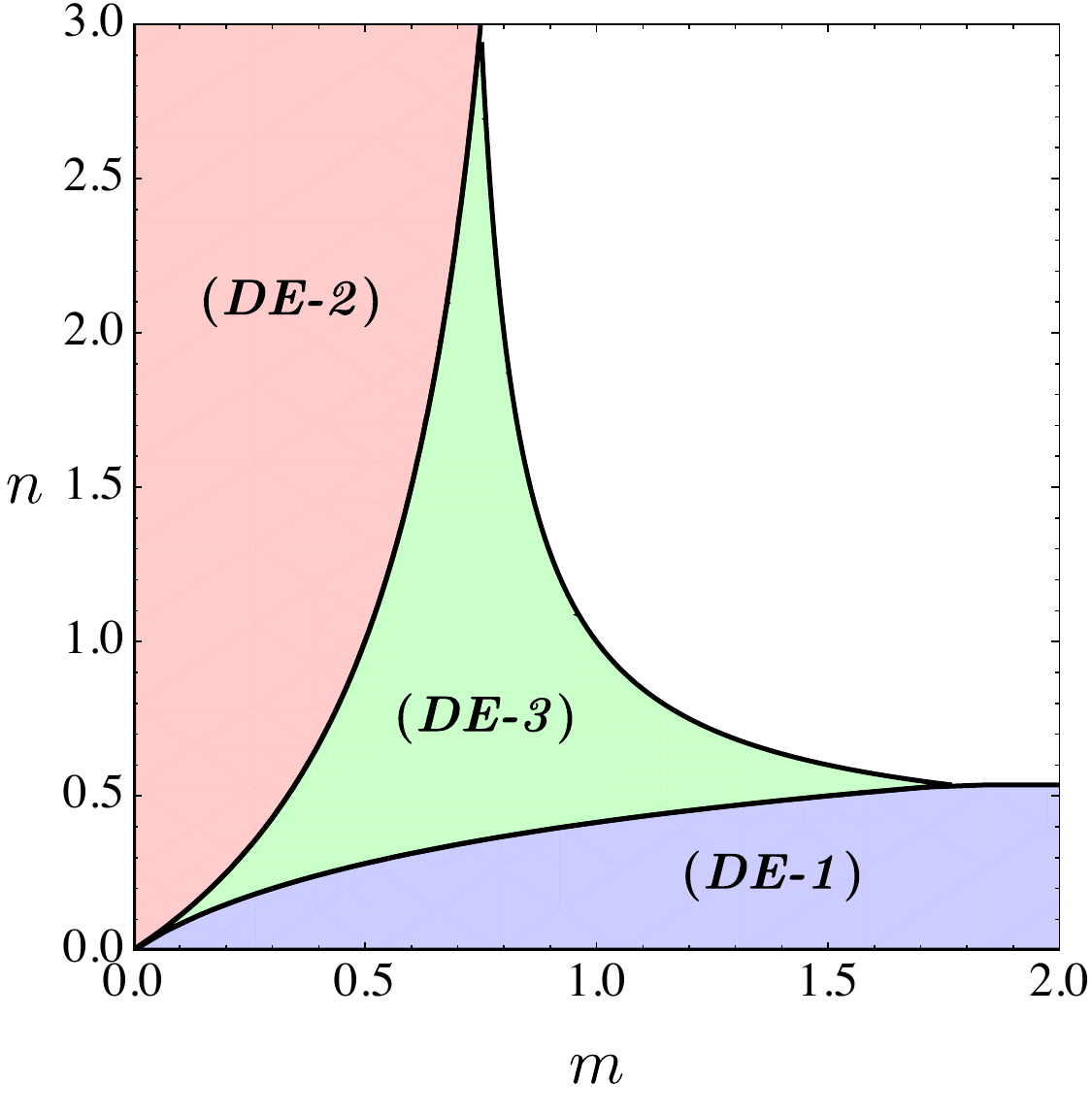}    
\caption{ Stability regions for the dark energy dominated points. Each color represents a $(m, n)$ parameter region where the indicated fixed point is an attractor. These three regions are separated by bifurcations curves (black solid lines), i.e. they are mutually excluded.}
\label{Regions}
\end{figure}

From the previous analysis, it is clear that the three dark energy dominated points can be attractors. In Fig. \ref{Regions}, we plot the $(m, n)$ parameter space where each dark energy dominated point is an attractor. We can see that these regions are separated by bifurcation curves (black solid lines), meaning that they are mutually excluded. This ensures that the system has only one dark energy attractor for a particular set of parameters $(m, n)$. This leave the case $n = m$ as the only possibility to get an isotropic Universe, since the points ($DE$-1) and ($DE$-2) are saddle while ($DE$-3) is the global attractor under this condition.

\section{Cosmological evolution} \label{CE}

Figure~\ref{Regions} summarizes our main results regarding the theoretical viability of the solid as a model of anisotropic dark energy. In this section we want to study the dynamics of the model for the parameters $(m, n)$ inside the coloured regions of Fig. \ref{Regions}. We will implicitly assume that, prior to the radiation epoch, the Universe underwent an inflationary period which perfectly smoothed any initial spatial shear or inhomogeneities. Therefore, we choose $\Sigma_i = 0$ as an initial condition\footnote{Here, the subscript $i$ means that the corresponding quantity is evaluated at some time  deep in the radiation epoch.}, such that the starting point for any cosmological trajectory is from the isotropic radiation point (\emph{R-1}). Moreover, we will choose parameters $m$ and $n$ such that the attractor point is given by (\emph{DE-2}). This choice allows us to give a simple and concrete example of the cosmological dynamics which can be easily extended to the other attractor points. Since $\Sigma_i = 0$, it is natural to assume that the contributions to the energy budget coming from the variables $f_1$ and $f_2$ are the same at the starting point. Having this in mind, we have chosen
\begin{equation} \label{Init Cond}
\Omega_{r_i} = 0.99995, \ f_{1_i} = f_{2_i} = 10^{-14}, \ \Sigma_i = 0
\end{equation}
as initial conditions at the redshift $z = 7.25 \times 10^7$, and we have integrated the system up to $z\rightarrow -1$. Since observations favor an equation of state of dark energy close to $-1$, from Eq. \eqref{DE2 wDE} we have to choose \mbox{$m \approx 0$}. In particular, we have chosen $m = 10^{-5}$. From Fig. \ref{Regions}, we can see that there are less restrictions regarding the choice of the parameter $n$. For example, we could assume a value for $n$ allowing the existence of the scaling points (\emph{R-2}) or (\emph{M-2}). However, for simplicity, we have chosen $n = 10^{-2}$, such that the scaling points do not exist.

\begin{figure}[t!]
\includegraphics[width=0.95\linewidth]{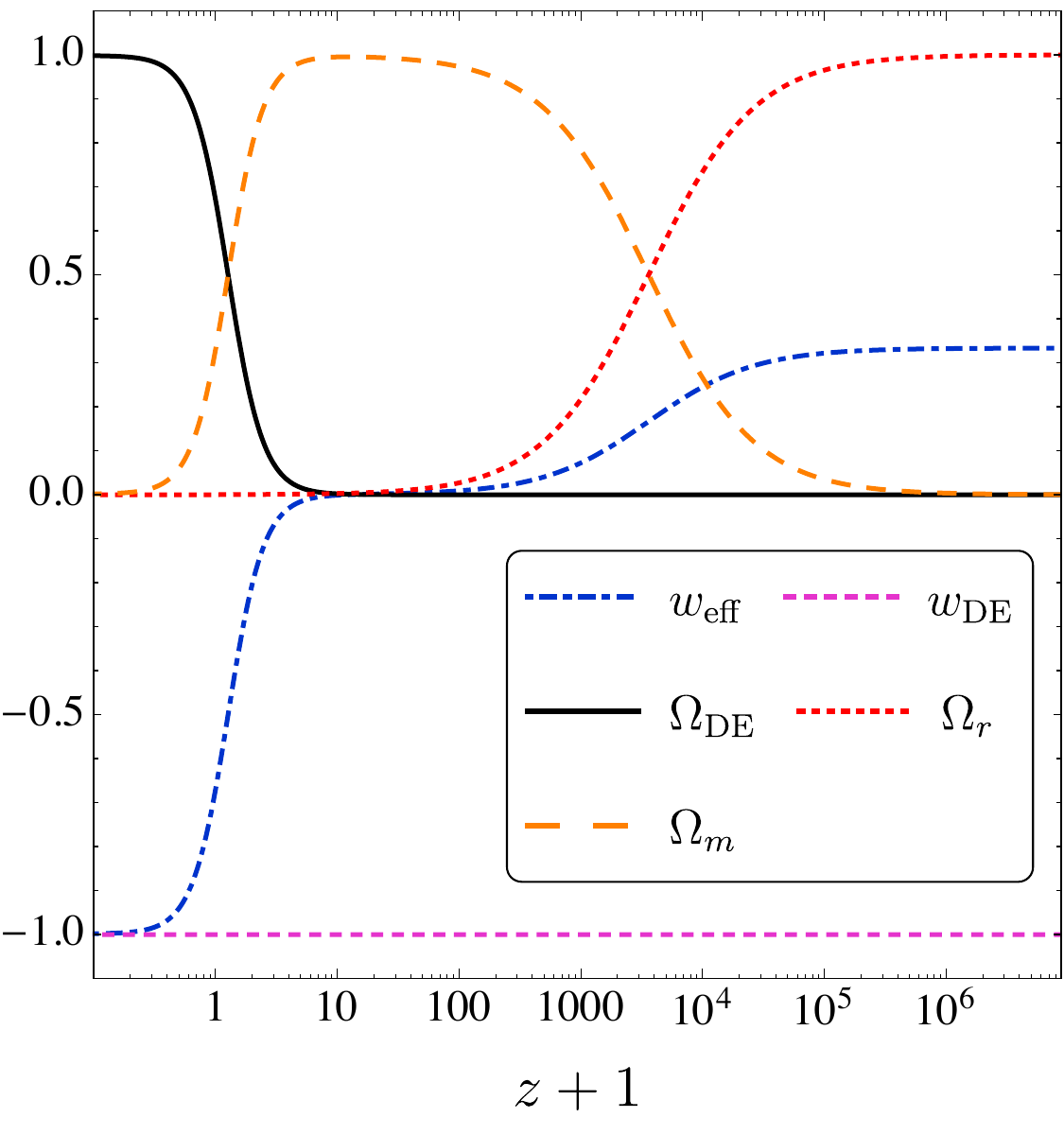}    
\caption{Evolution of the density parameters, the effective equation of state, and the equation of state of dark energy during the whole expansion history. The initial conditions were chosen deep in the radiation era at the redshift $z = 7.25 \times 10^7$. The Universe passes through radiation dominance at early times (red dotted line), followed by a matter dominance (light brown dashed line), and ends in the dark energy dominance (black solid line) characterized by $w_\text{eff} \simeq -1$ (blued dot-dashed line). The dark sector behaves very similar to a cosmological constant since $w_\text{DE} \simeq -1$ (magenta small-dashed line).}
\label{AniAbundances}
\end{figure}

In Fig. \ref{AniAbundances}, we plot the dynamical evolution of $\Omega_r$, $\Omega_m$, $\Omega_{\text{DE}}$,  $w_{\text{eff}}$ and $w_{\text{DE}}$ obtained from the numerical integration of Eqs. (\ref{f1 eq})-(\ref{r eq}). In the case where $\Sigma_i \approx 0$, there are no appreciable changes in the cosmological behavior presented in Fig. \ref{AniAbundances}. However, a ``stiff matter'' epoch driven by the spatial shear appears before the radiation era, which we briefly treat in Appendix \ref{Other fixed point}.

In particular, Fig. \ref{AniAbundances} shows that the radiation-dominated epoch ($\Omega_r \approx 1$ and $w_{\text{eff}} \approx 1/3$) runs from $z = 7.25 \times 10^7$ to $z \approx 3200$ where the radiation-matter transition occurs. Moreover, $\Omega_{\text{DE}} \approx 3.52 \times 10^{-11}$  during this transition, obeying the BBN constraint $\Omega_{\text{DE}} < 0.045$ \cite{Bean:2001wt}. The length of this radiation phase is in agreement with the constraint given in Ref. \cite{Alvarez:2019ues}. From $z \approx 3200$, the Universe is dust-dominated  ($\Omega_m \approx 1$ and $w_{\text{eff}} \approx 0$) until $z \approx 0.3$ at the matter-dark energy transition. The contribution of the dark sector is $\Omega_\text{DE} \approx 1.58 \times 10^{-5}$ at $z = 50$, value which is within the CMB bound $\Omega_\text{DE} <  0.02$ \cite{Ade:2015rim}. The dark energy-dominance ($\Omega_\text{DE} \approx 1$ and $w_{\text{eff}} < -1/3$)  starts from $z = 0.3$ and on into the future, agreeing with the results given by the dynamical system analysis, i.e. (\emph{DE-2}) is an attractor. This is further supported by the fact that the values of $f_1, f_2, \Sigma$ and $w_\text{DE}$ are those predicted by the dynamical system. Explicitly, $f_1 \approx 0$, $f_2 \approx 0.707107$, $\Sigma \approx 3.33334 \times 10^{-6}$, and $w_\text{DE} \approx - 0.999993$ in the far future ($z \rightarrow -1$), which are consistent with the values computed from the (\emph{DE-2}) line in Table~\ref{table-fp} and Eq.~(\ref{DE2 wDE}). Although $w_\text{DE}$ seems to be constant during the whole expansion history, this is not the case. Indeed, during the radiation-matter transition ($z \approx 3200$) we have $w_\text{DE} \approx - 0.998059$, while during the matter-dark energy transition ($z \approx 0.3$) we have $w_\text{DE} \approx - 0.998171$. The final value is $w_\text{DE} \approx - 0.999993$, which corresponds to the value in the attractor point [see Eq. (\ref{DE2 wDE})]. Thus, the numerical solution shows a nearly constant varying equation of state of the dark energy, changing only about $\sim 0.001 \%$ during this particular cosmological trajectory. This tiny variation in $w_\text{DE}$ is impossible to verify by current technology, since it is well below the threshold of missions like Planck or Euclid \cite{Aghanim:2018eyx, Laureijs:2011gra}.

From Fig. \ref{AniAbundances}, we can also notice that dark energy does not behave as radiation or dust, given that $w_\text{DE} \neq \{1/3, 0\}$, confirming that the Universe does not approach the scaling anisotropic points. Indeed, we have confirmed, for several pairs of parameters $(n, m)$, that the expansion history of the Universe is very similar to that shown in Fig. \ref{AniAbundances}, and thus the cosmological trajectories of the Universe are never close to the scaling anisotropic points. We can thus conclude that the typical cosmological evolution is given by
$$(\emph{R-1}) \to (\emph{M-1}) \to (\emph{DE-i}),$$
where $i$ can be 1, 2 or 3, depending on the $m$ and $n$ values we choose (see Fig. \ref{Regions}).

\begin{figure}[t!]
\includegraphics[width=0.85\linewidth]{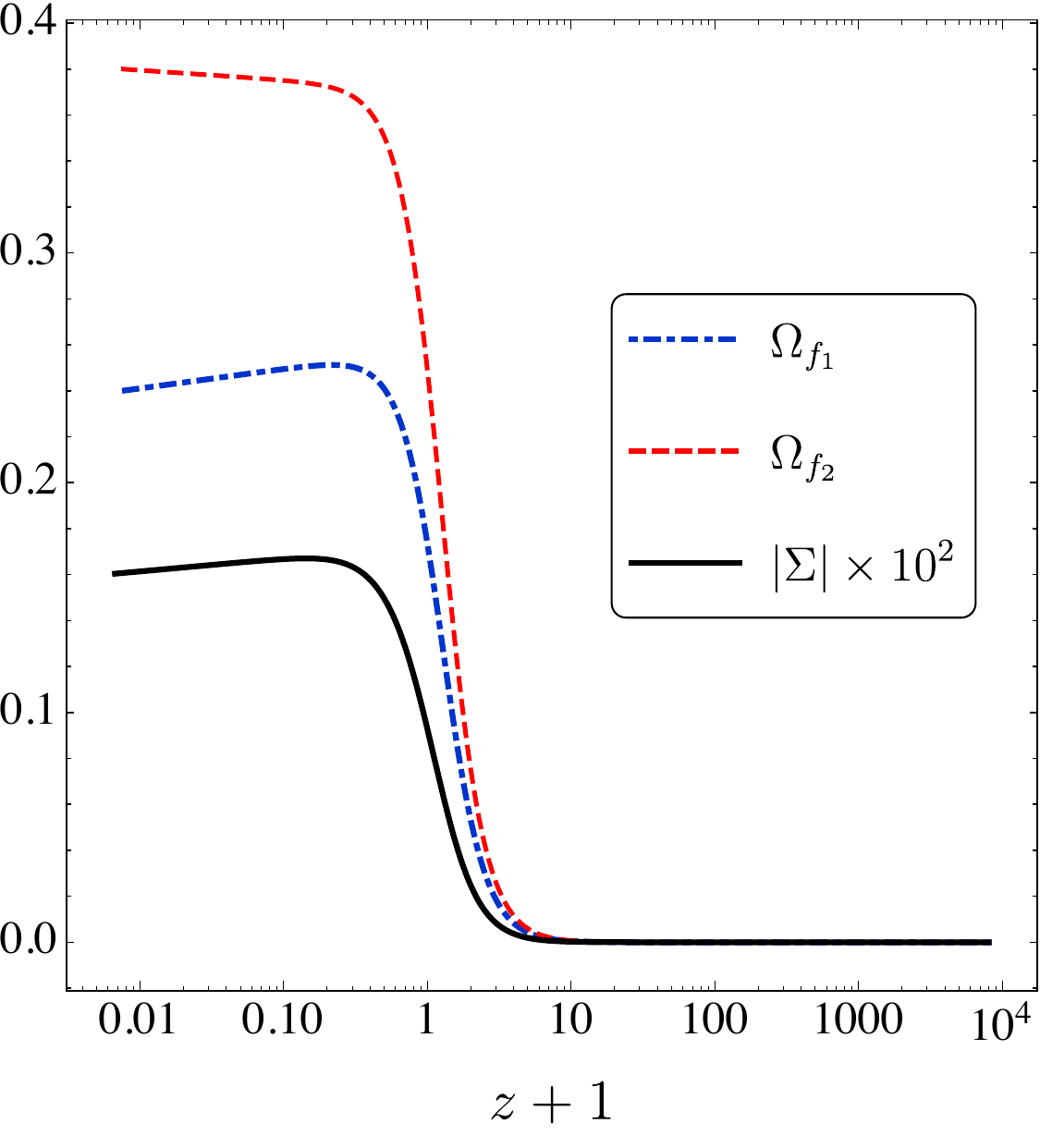}    
\caption{Evolution of the density parameters associated to the solid Lagrangians and the shear for the initial conditions in Eq. \eqref{Init Cond} and parameters $m = 10^{-5}$ and $n = 10^{-2}$, such that (\emph{DE-2}) is the attractor point.}
\label{DEComponents}
\end{figure}

As shown in Fig. \ref{DEComponents}, The density parameters $f_1^2 \equiv \Omega_{f_1}$ and $f_2^2 \equiv \Omega_{f_2}$ associated with each solid Lagrangian $F^I$ grow during the late matter-dominated epoch around $z = 10$, while they are subdominant in the whole prior cosmological evolution, as expected from the dynamical analysis. 

\begin{figure}[t!]
\includegraphics[width=0.95\linewidth]{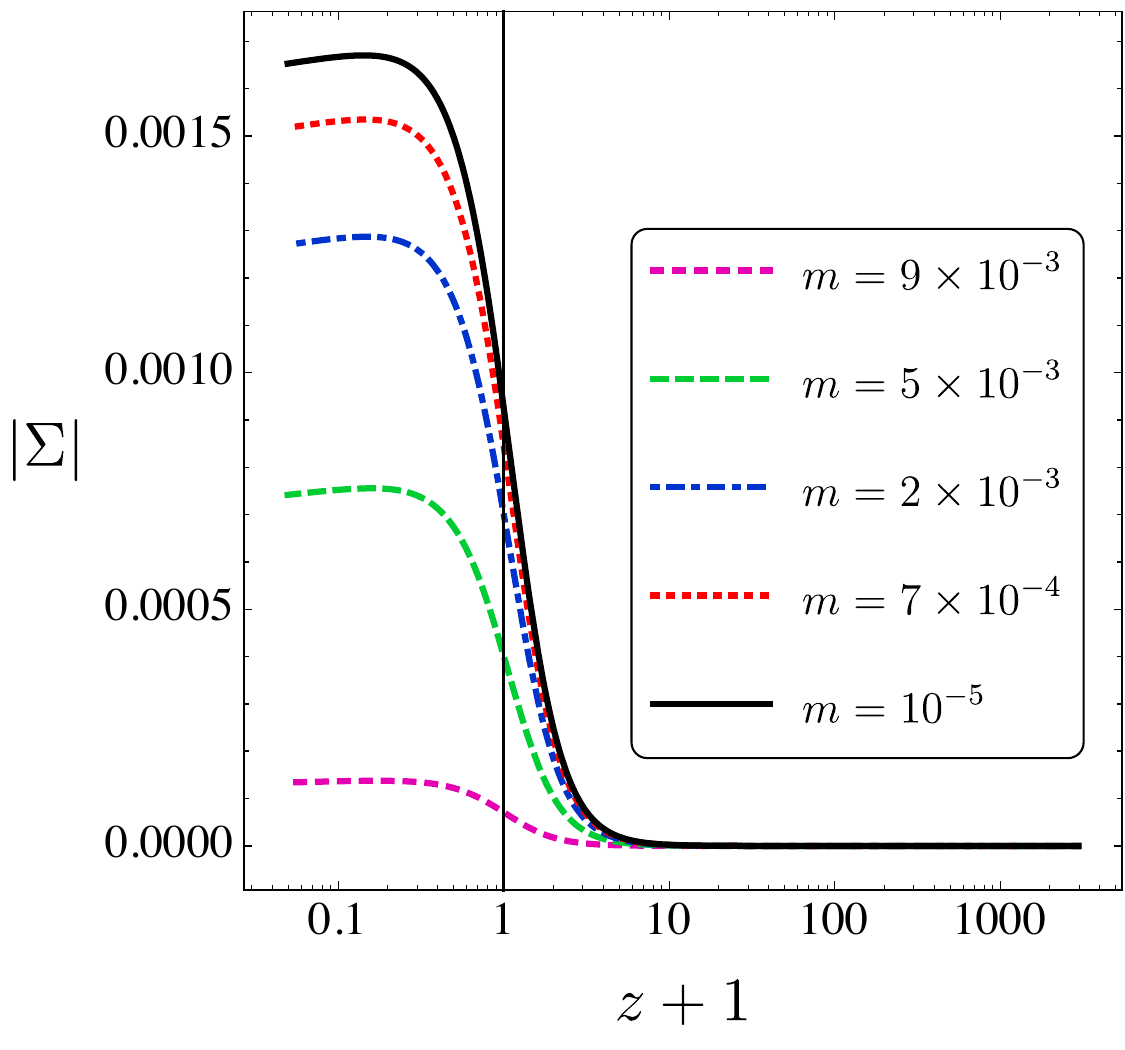}    
\caption{(Color online) Evolution of the shear $\Sigma$ around $z = 0$ for different values of the parameter $m$, while $n$ is fixed and the initial conditions are the same given in Eq. \eqref{Init Cond}. The thin vertical line signals the value of the shear today.}
\label{Shear}
\end{figure}

We have also investigated the evolution of $\Sigma$ taking the same initial conditions in Eq. (\ref{Init Cond}) and $n = 10^{-2}$, but this time with different choices for the parameter $m$ in such a way that (\emph{DE-2}) is the only attractor of the system. The results are shown in Fig. \ref{Shear}, where we can see that $| \Sigma |$ starts to grow around $z = 10$, similarly to $f_1$ and $f_2$ as seen in Fig. \ref{DEComponents}. The values predicted for the present spatial shear are $| \Sigma_0 | \leq 9.4 \times 10^{-4}$, corresponding to the black solid curve in Fig. \ref{Shear}, which are in agreement with the observational bounds $| \Sigma_0 | \leq \mathcal{O} (0.001)$ \cite{Campanelli:2010zx, Amirhashchi:2018nxl}. We want to stress that, even for $\Sigma_i = 0$, the final state of the Universe is an anisotropic accelerated expansion.

On the other hand, we corroborated that the other two dark energy points, namely (\emph{DE-1}) and (\emph{DE-3}), can be attractors in their respective stability regions (see Fig. \ref{Regions}). In order for (\emph{DE-1}) to be the only attractor of the system, we choose $n = 10^{-5}$ and $m = 10^{-3}$ according to the existence and stability conditions in Eqs.  \eqref{DE1 region} and \eqref{DE1 attractor}. The initial conditions are the same as in Eq. \eqref{Init Cond} and, although the values of $f_1, f_2, \Sigma$ and $w_\text{DE}$ change, the cosmological evolution is qualitatively the same. In particular, the shear and the equation of state of dark energy today are $| \Sigma_0 | \approx 1.2 \times 10^{-4}$ and $w_{\text{DE}_0} \approx -0.999559$. We confirm that the values predicted by the dynamical system analysis are in agreement with our  numerical solution: $f_1 = 1, f_2 = 0, \Sigma = -6.66669\times10^{-6}$ and $w_{\text{DE}} = -0.999993$ [see Table~\ref{table-fp} and Eq.~(\ref{FixedDE1wDE})], when evaluated for redshifts $z \rightarrow -1$. Finally, for the point (\emph{DE-3}), we choose $n = 10^{-3}$ and $m = 10^{-3}$, so that this point is the only attractor of the system. Using the same initial conditions as in Eq. (\ref{Init Cond}), the cosmological dynamics is not significantly changed apart from the important fact that $\Sigma = 0$ during the whole cosmological evolution, i.e. the expansion history is isotropic. By changing the values of the parameters $n$ and $m$ while keeping (\emph{DE-3}) as the only attractor, we verified that the case $n = m$ is the unique isotropic solution as expected from the discussion in Sec. \ref{SecDE3}. The numerical results are also consistent with the analytical findings, namely: $f_1 = f_2 = 0.57735, w_{\text{DE}} = -0.999333$ and $\Sigma = 0$ for redshifts $z \lesssim -0.99$ [see Table~\ref{table-fp} and Eq. (\ref{FixedDE3wDE})], i.e. the attractor point is reached very quickly.

\subsection{Observational Signatures}

From the magnitude-redshift data of SNe Ia, the analysis of Ref. \cite{Campanelli:2010zx} showed that the present value of the spatial shear is constrained to be $| \Sigma_0 | \lesssim \mathcal{O}(0.01)$. More recently, the authors of Ref. \cite{Amirhashchi:2018nxl} used a combination of the observational Hubble data $H(z)$ and SNe Ia measurements to put the tighter bound of $| \Sigma_0 | \lesssim \mathcal{O}(0.001)$. Similarly, future weak-lensing measurements with the Euclid satellite are expected to reach a similar sensitivity level of $|\Sigma_0|\lesssim{\cal{O}}(0.008)$~\cite{Pereira:2015jya}. For the set of parameters and initial conditions used in Fig. \ref{Shear}, we can see that $| \Sigma_0 |$ is within these bounds, although it can take greater values in the future cosmological evolution. 

An anisotropic dark energy has the potential for breaking of statistical isotropy and explain of some of the large scale CMB anomalies. In particular, anisotropic pressure can induce peculiar velocity flows, anisotropy in the SNe Ia data, and significant CMB dipole and quadrupole \cite{Perivolaropoulos:2014lua} fluctuations. The main observational effect of an anisotropic shear is its contribution to the CMB temperature anisotropies. This contribution is introduced through the redshift at last scattering surface, which becomes anisotropic. Considering only large-scale fluctuations, this can be quantified as \cite{Appleby:2009za, Almeida:2019iqp}:
\begin{equation}
\left| \frac{\delta T}{T} (\hat{\boldsymbol{n}}) \right| \lsim \left| \sigma_0 - \sigma_\text{dec} \right|,
\end{equation}
where $\delta T / T$ is the CMB temperature anisotropies, $\hat{\boldsymbol{n}}$ is the unit vector along the line-of-sight, and $\sigma_\text{dec}$ is the value of the geometrical shear at the time of decoupling ($z \simeq 1090$). Since $\sigma' = \Sigma$, we can obtain the values for the geometrical shear by  numerical integration. For the cases with the largest and smallest shear today, as depicted in Fig. \ref{Shear}, and using the same initial conditions of Eq. \eqref{Init Cond}, we obtain
\begin{align}
\text{largest} \ \Sigma_0: \ \left| \sigma_0 - \sigma_\text{dec} \right| \approx 5 \times 10^{-4}, \\
\text{smallest} \ \Sigma_0: \ \left| \sigma_0 - \sigma_\text{dec} \right| \approx 3 \times 10^{-5},
\end{align}
which are in qualitative agreement with the conservative bound $\left| \sigma_0 - \sigma_\text{dec} \right| < 10^{-4}$ \cite{Aghanim:2018eyx}, thus, a solid anisotropic dark energy could alleviate the observed CMB quadrupole anomaly.

\section{Isotropic Dark Energy} \label{Iso DE}

So far, we have been mainly interested in the solid as a viable model for an anisotropic late-time expansion of the Universe. Interestingly, in the previous analysis, we showed that the only possibility to get an isotropic Universe is the case when $n = m$. 
Nonetheless, the solid is also compatible with the FLRW metric, and it can also be used to describe an isotropic late-time expansion with equation of state close to $-1$. For this we have to consider that the Lagrangians $F^I$ are identical. Thus, the isotropic model requires $\sigma = 0$ in the Bianchi-I metric in Eq. \eqref{metric}, and:
\begin{equation}
X^I = X = 1 / a^2, \quad F^I = F, \quad I = 1, 2, 3. 
\end{equation}

The Friedman equations in Eqs. (\ref{H2 eq}) and (\ref{Hdot eq}) are simplyfied as 
\begin{align} 
3 \, m_\text{P}^2 H^2 & = 3 F + \rho_m + \rho_r,\label{Iso H2 eq} \\
- 2 \, m_\text{P}^2 \dot{H} & = 2 X F_X + \rho_m + \frac{4}{3}\rho_r. \label{Iso Hdot eq} 
\end{align}

Defining $f^2 = F / (m_\text{P}^2 H^2)$ and using the density parameter $\Omega_r$ in Eq. \eqref{variables}, the autonomous set is reduced to
\begin{align} 
f' &= f \left(q + 1 - x \right)\,, \label{Iso f1 eq}\\
\Omega_r' &= 2 \, \Omega_r (q - 1), \label{Iso r eq}
\end{align}
where $\Omega_m$ is given in terms of $f$ and $\Omega_r$ by the Friedman constraint coming from Eq. (\ref{Iso H2 eq}), and the deceleration parameter is given by
\begin{equation} \label{deceleration2}
q = \frac{1}{2} \left[ 1 + (2 x - 3) f^2 + \Omega_r \right],
\end{equation}
where $x = X F_X / F$ is a function characterizing the form of the Lagrangian $F$. Assuming a power law model, $F \propto X^n$, we get $x = n$ and the system is closed. Three fixed points can be found: a radiation dominated point which is a source for $n < 2$, a matter dominated point which is a saddle for $n < 3/2$, and a dark energy dominated point which is an attractor for $n < 3/2$.

\begin{figure}[t!]
\includegraphics[width=0.95\linewidth]{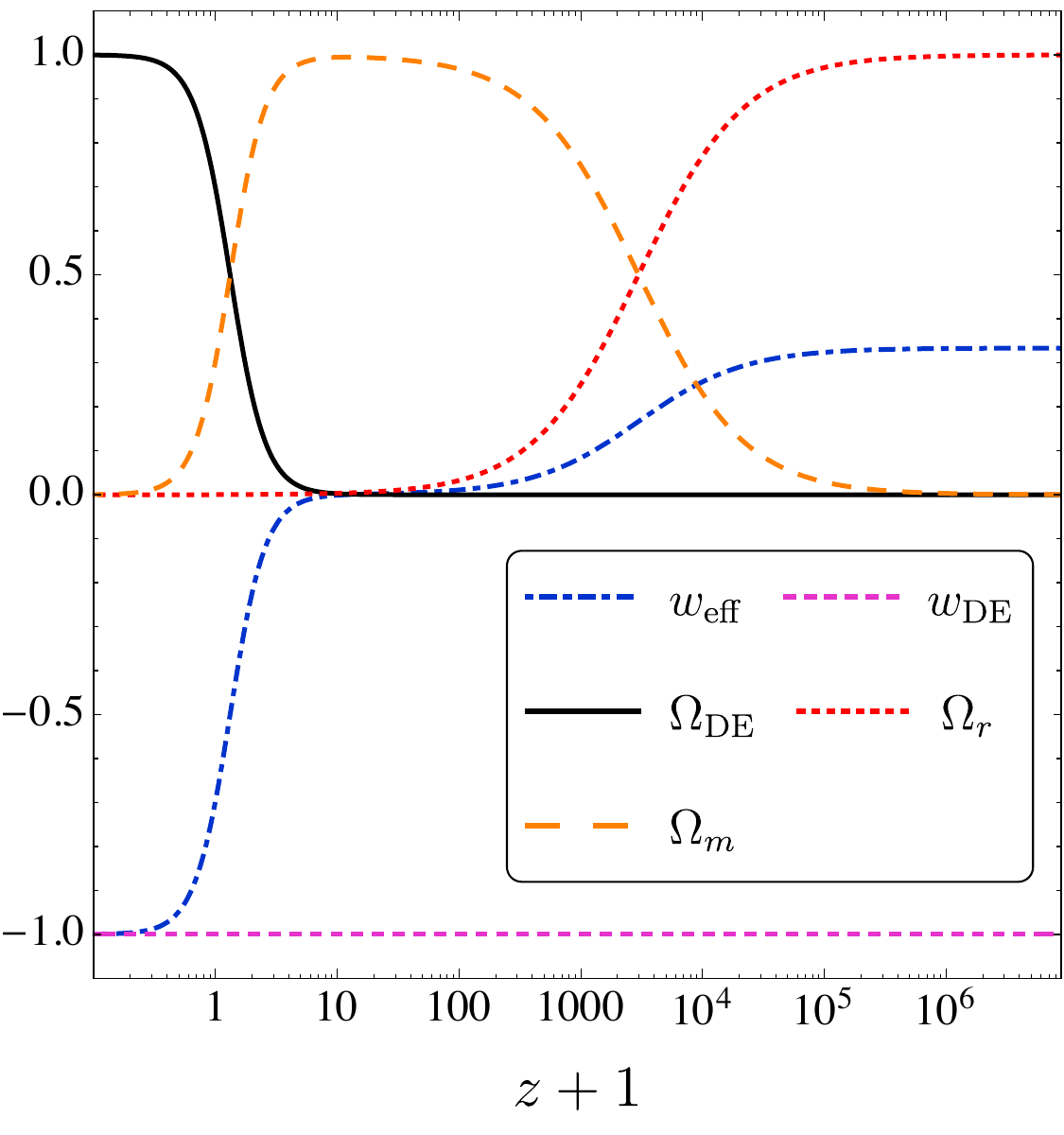}    
\caption{Evolution of the density parameters, the effective equation of state, and the equation of state of dark energy for $n=10^{-3}$. The initial conditions were chosen according with the present observed values for the density parameters, {\it i.e.}: $\Omega_{r, 0} = 10^{-4}, \ \Omega_{m, 0} = 0.3, \ \Omega_{\text{DE}_0} \equiv f_0^2 \approx 0.7.$  The Universe begins in a radiation dominated epoch (red dotted line), followed by a matter dominance one (light brown dashed line) and ends in the dark energy epoch (black solid line) characterized by $w_\text{eff} \simeq -1$ (blued dot-dashed line). We can see that the dark sector behaves as a cosmological constant since $w_\text{DE} \simeq -1$ (magenta small-dashed line).}
\label{AbundancesIC}
\end{figure}

In Fig. \ref{AbundancesIC}, we numerically integrate the set given by Eqs. (\ref{Iso f1 eq}) and (\ref{Iso r eq}). We assume $n = 10^{-3}$ and the initial conditions according with the present observed values for the density parameters: \cite{Aghanim:2018eyx}:
\begin{equation} 
\Omega_{r, 0} = 10^{-4}, \ \Omega_{m, 0} = 0.3, \ \Omega_{\text{DE}_0} \equiv f_0^2 \approx 0.7. 
\end{equation}
We can see that the cosmological evolution is quite similar to the one of the anisotropic case, which is shown in Fig. \ref{AniAbundances} and detailed in Sec. \ref{CE}.

In contrast to the quintessence model, this model does not have a kination epoch previous to the radiation domination period. This is due to the fact that the kination epoch requires $\dot{\phi}^2 \gg V(\phi)$, where $\phi$ is the quintessence scalar field and $V$ its potential. In the quintessence model, the potential is necessary in order to get accelerated expansion since this stage is provided by $\dot{\phi}^2 \ll V(\phi)$. In the present case, such potential is not needed since the accelerated expansion is driven by the proper kinetic terms of the scalar fields through the function $F$. Another important difference between this model and quintessence is that the equation of state of dark energy can be constant or dynamical. The equation of state of dark energy  is
\begin{equation} \label{Iso wDE}
w_{\text{DE}} = - 1 + \frac{2}{3} x,
\end{equation}
which is constant for constant $x$ (power law model), thus mimicking a cosmological constant. However,  $w_\text{DE}$ can also be dynamical for a general time-dependent $x$ function. 

\section{Conclusions} \label{conclusions}

In this work we studied a set of three scalar fields with a constant but nonvanishing spatial gradients as a source of anisotropic dark energy. This particular set of inhomogeneous scalar fields, known as ``solid'', has been used in the inflationary context showing interesting features at the perturbative level \cite{ArmendarizPicon:2007nr, Endlich:2012pz}. It was also shown that anisotropic inflationary solutions are possible \cite{Bartolo:2013msa, Bartolo:2014xfa}. Here, we have investigated the late-time dynamics of the solid in a Bianchi-I expanding universe. Through a dynamical system analysis, we showed that for a particular power-law model, the solid can generate an anisotropic late-time accelerated expansion, being this epoch an attractor of the system for suitable values of the free parameters of the model (see Fig. \ref{Regions}). In these regions, the cosmological evolution starts in a radiation-dominance epoch, followed by a matter-dominance epoch, and ending in a possible anisotropic dark energy-dominated period which can be realized by three different points, whose attractor regions are separated by bifurcation curves, i.e. they are mutually excluded. Nonetheless, we found that in the particular case when $n = m$ the Universe isotropizes provided that the Lagrangians $F^I$ evolve in the same way.

The dynamical analysis was complemented with a numerical integration of the dynamical system. The parameters were fixed in such a way that (\emph{DE-2}) was the only attractor point and the equation of state of dark energy was close to $-1$ as expected from observations \cite{Aghanim:2018eyx}. The initial conditions were chosen in the deep radiation era (redshift $z = 7.25 \times 10^7$) and assuming a zero spatial shear ($\Sigma_i = 0$). We found that the spatial shear differs sensibly from zero at around $z = 10$, taking nonnegligible values nowadays but within the observational bounds $| \Sigma_0 | \lesssim \mathcal{O} (0.001)$ \cite{Campanelli:2010zx, Amirhashchi:2018nxl}. The numerical solution also showed a nearly constant equation of state of dark energy, $w_\text{DE}$, that only changed about $\sim 0.001 \%$ from its value at $z = 7.25 \times 10^7$ to its final value in the far future ($z \rightarrow -1 $). We verified that similar behaviors are obtained if the parameters are chosen to establish (\emph{DE-1}) or (\emph{DE-3}) as the attractor points. 

The nonvanishing shear after the radiation-dominated epoch leaves imprints on CMB and SNe  Ia data. In particular, the shear affects the CMB quadrupole temperature anisotropy through the standard Sachs-Wolfe formula.  We showed that the change of the spatial shear from decoupling to today can be compatible with the CMB quadrupole data. In particular, if $\left| \sigma_0 - \sigma_\text{dec} \right|$ is of order $10^{-4}$, there may be an interesting possibility for addressing the CMB quadrupole anomaly.

We have also investigated the isotropic version of the solid model as a candidate for (isotropic) dark energy. In this case, only three fixed points were found: a source radiation point, a saddle matter point, and an attractor dark energy point. The evolution of the Universe is very similar to the anisotropic case (see Figs. \ref{AniAbundances}). However, for a power-law model, the equation of state of dark energy is indeed a constant [see Eq. (\ref{Iso wDE})], and is thus phenomenologically distinguishable from quintessential models.

\section*{Acknowledgements}
This work was supported by the following grants: Vicerrector\'ia de Investigaciones $-$ Universidad del Valle Grant No. 71220 and Patrimonio Autónomo - Fondo Nacional de Financiamiento  para  la  Ciencia,  la  Tecnología  y  la  Innovación  Francisco  José  de  Caldas  (MINCIENCIAS - COLOMBIA)  Grant  No.   110685269447  RC-80740-465-2020,  project  69723 T.S.P thanks Brazilian funding agencies CAPES and CNPq (grants 438689/2018-6 and 311527/2018-3) for the financial support. 
\appendix

\section{Other fixed point} \label{Other fixed point}
The fixed points presented are the relevant point for the cosmological history of the Universe. Nonetheless, the dynamical system has another fixed points, which we present here.

\paragraph{($S$) Stiff matter domination:} This fixed point is characterized by
\begin{equation}
f_1 = 0, \ f_2 = 0, \ \Sigma = \pm 1, \ \Omega_r = 0,
\end{equation}
with $\Omega_m = 0$, $\Omega_{\text{DE}} = 1$ and $w_{\text{DE}} = 1$. Since $w_\text{eff} = 1$, these points correspond to a ``stiff matter'' domination driven by the spatial shear $\Sigma$. This also implies that the energy density of dark energy decays as $\rho_{\text{DE}} \propto a^{-6}$, and thus this period is prior to the radiation domination. Although these points are the only possible sources of the model (i.e., all the eigenvalues of the Jacobian matrix $\mathbb{M}$ are positive), they can be arbitrarily pushed back to the past depending on how small $\Sigma_i$ is. For example, these points are in the infinite past for $\Sigma_i = 0$. This stiff fluid is  common in quintessence models where the kinetic term of the scalar field has the chance to dominate \cite{Amendola:2015ksp}, and it has been pointed out in some works that this period can be useful in the study of the reheating process \cite{Ferreira:1997hj, Pallis:2005bb, Dimopoulos:2018wfg, Bettoni:2018pbl, Bettoni:2018utf, Bettoni:2019dcw}. 

\bibliography{Bibli.bib}

\end{document}